\documentclass[12pt,preprint]{aastex}

\usepackage{color}

\begin{document}

\title{A cgi synthetic CMD calculator for the YY Isochrones}
\author{P. Demarque\altaffilmark{1}, S. N. Virani\altaffilmark{1}, E. J. Murphy\altaffilmark{1,2}, J.-H. Woo\altaffilmark{1,3}, Y. -C. Kim\altaffilmark{4} \& S. K. Yi\altaffilmark{4}}

\affil{$^1$ Department of Astronomy, Yale University, New Haven, CT 06520-8101, USA}

\affil{$^2$ Spitzer Science Center, California Institute of Technology, Pasadena, CA 91125, USA}

\affil{$^3$ Department of Physics, University of California, Santa Barbara, CA 93106-9530, USA}

\affil{$^4$ Department of Astronomy, Yonsei University, Seoul 120-749, Korea}

\begin{abstract}
We describe a web-based cgi calculator for constructing synthetic  
color-magnitude diagrams for a simple stellar population (SSP) using the 
Yonsei-Yale (YY) isochrone data base.  
This calculator is designed to be used interactively.  It creates quick look 
CMD displays in (B-V) and (V-I) colors.  Stochastic effects on the CMDs 
are included.  Output in tabular form is 
also provided for special purpose displays, or for combining 
the CMDs of different stellar populations.
This research tool has applications in studies  
of the stellar content of our Galaxy and external systems.  It  provides 
an easy way to interpret the CMDs in resolved 
stellar populations.  It offers the means to explore the dependence 
of the integrated properties of unresolved stellar systems on 
stellar parameters (ages, chemical composition, binarity) and on 
the characteristics of their parent population (IMF slope and 
mass range).
\end{abstract}

\section{Introduction}

This program creates a synthetic color-magnitude diagram (CMD)
for a simple stellar population (SSP), i.e. for a stellar population of a 
given age and chemical composition, that obeys a specified initial mass 
function (IMF) based on the prescription by Park \& Lee (1997).  
Stochastic effects on the CMD are included.
As an additional feature, the presence of a population of unresolved binaries can 
be taken into account (Woo et al. 2003).

The code is designed for use with the Yonsei-Yale (YY) isochrones, and covers the pre-main sequence and hydrogen burning phases of evolution.  The YY isochrones and luminosity functions are described in detail in papers by Yi et al. (2001, 2003; Paper~1 and Paper~3) and Kim et al. (2002; Paper~2).  Evolutionary tracks are available in Demarque et al. (2004; Paper~4).  

The YY isochrone tables cover metallicities $Z$ (heavy element content by mass) in the range 0.0 to 0.08, and values of $\alpha$ enhancement corresponding to  $[\alpha/Fe]$ = 0.0, 0.3 and 0.6.
Helium diffusion and convective core overshoot have been taken into account in calculating the evolutionary tracks.  This first set of isochrones (Paper ~1) was for the scaled solar mixture. The completing sets (for twice and four-times $\alpha$-enhanced) were released in Paper~2. Two significant features of these isochrones are that (1) the stellar models start their evolution from the pre-main sequence birthline instead of from the zero-age main sequence, and (2) the color transformation has been performed using both the tables of Lejeune et al. (1998), and the older, but now modified, Green et al. tables (1987).

\begin{deluxetable}{lccc}
\tablewidth{0pt}
\tablecaption{The assumptions for $\alpha$-enhancement
 \label{tbl1}}
\tablehead{\colhead{Element} &\colhead{[$\alpha$/Fe]=0.0}\tablenotemark{a} &
\colhead{[$\alpha$/Fe]=+0.3} &\colhead{[$\alpha$/Fe]=+0.6} }
\startdata
C & 8.55 &  & \\
N & 7.97 &  & \\
O & 8.87 & 9.17 &9.47 \\
Ne & 8.08 & 8.38 &8.68 \\
Na & 6.33 & 6.63 &6.93 \\
Mg & 7.58 & 7.88 &8.18 \\
Al & 6.47 & 6.17 &5.87 \\
Si & 7.55 & 7.85 &8.15 \\
P & 5.45 & 5.75 &6.05 \\
S & 7.21 & 7.51 &7.81 \\
Cl & 5.50 & 5.80 &6.10 \\
Ar & 6.52 & 6.82 &7.12 \\
K & 5.12 &  & \\
Ca & 6.36 & 6.66 &6.96 \\
Ti & 5.02 & 5.32 &5.62 \\
Cr & 5.67 &  & \\
Mn & 5.39 & 5.24 &5.09 \\
Fe & 7.50 &  & \\
Ni & 6.25 &  & \\
\enddata
\tablenotetext{a}{ The scaled-solar abundance ratios of metals are taken
 from Grevesse \& Noels 1993.}
\tablecomments{The abundance of the elements in logarithmic scale,
$\log N_{el}/N_H +12$, where $N_{el}$ is the abundance by number. }
\end{deluxetable}

The web interface to the CMD calculator can be found at the two YY web sites: 
\begin{center}
\underline{\textrm{\textcolor{blue}{http://www.astro.yale.edu/demarque/yyiso.html}}}\\or\\
\underline{\textrm{\textcolor{blue}{http://csaweb.yonsei.ac.kr/~kim/yyiso.html}}}\\
\end{center}
In addition, these two web sites provide links to the four YY  papers listed above, and codes to interpolate between isochrones in age and chemical composition within the available parameter range.

\section{CMD calculator input parameters}
A number of input parameters must be defined in the construction of synthetic CMDs. 
The list of input parameters on the web interface is given below.  The output CMD can be retrieved  
either in tabular or graphical form.
\begin{itemize}
\item{$\bf{Alpha}$}\\
This is the composition parameter  $[\alpha/Fe]$, which denotes the relative abundance by number with respect to the Sun of $\alpha$ enriched nuclei in the chemical composition mixture.  The notation is the usual logarithmic relative abundance with respect to the Sun.  In this notation, the metallicity of a star is defined by
\begin{equation}
 [X/H] \equiv log(N_{X}/N_{H})_{\star} - log(N_{X}/N_{H})_{\odot}
\end{equation}
where $N_{X}$ and $N_{H}$ are the relative abundances by numbers of element X and of hydrogen, respectively.  The abundance ratio $[\alpha/Fe]$ is defined correspondingly.  
Input values of $\alpha$ must be in the range 0.0 to 0.6 (by definition,  $\alpha$ = 0 corresponds to the solar mixture).
\item{$\bf{Metallicity}$}\\
The metallicity parameter $Z$ for the synthetic CMD.  $Z$  is the mass fraction of the heavy elements as usually defined in stellar interior calculations (i.e. X + Y+ Z = 1, where X and Y are the mass fractions of H and He).  The parameter $Z$ should be in the range (near) 0.0 to 0.08 (Note that $Z \equiv 0$ is not allowed).  Table~2 provides a conversion between $Z$ and $[Fe/H]$ for different values of  $[\alpha/Fe]$.
\item{$\bf{Age}$}\\
The age of the stellar population (in Gyr) for the calculated synthetic CMD. 
\item{$\bf{Number~of~stars}$}\\
The total number of stars $N$ to be used in the calculation of the synthetic CMD.  A maximum of 100,000 stars is allowed for the value of  N. 

The total number of stars $N$ in the synthetic CMD is defined as:
\begin{equation}
  N = C \int_{M_{lower}}^{M_{upper}} M^{-(x+1)}\, dM
 \end {equation}
where $x$ is the IMF slope, and $M_{upper}$ and $M_{lower}$ define the range of masses included in the integration.  The definition of $x$ is the same as in the review by Tinsley (1980).  The YY luminosity functions are tabulated for $x$ = -1, 1.35 and 3.  The $x$ = 1.35 power law corresponds to the Salpeter IMF (1955).  C is a normalization constant.  The YY tables are normalized to 1000 stars in the mass range 0.5-1.0 $M_{\odot}$.    

\item{$\bf{IMF~slope}$}\\
Chosen value of the IMF slope $\bf{x}$ defined in eq. (2) (Note that $\bf{x} \neq 0$).
\item{$\bf{IMF~upper~limit}$}\\
The upper mass limit $M_{upper}$ of the IMF in eq.(2), in solar units.
\item{$\bf{IMF~lower~limit}$}\\
The lower mass limit $M_{lower}$ of the IMF in eq.(2), in solar units.
\item{$\bf{ISEED}$}\\
The random number generator seed to be used for the synthetic CMD calculations.  Varying ISEED illustrates the importance of stochastic effects in CMDs for a given set of input parameters.  A knowledge of ISEED allows the user to retrieve previously made synthetic CMDs.
\item{$\bf{Binary~fraction}$}\\
The fraction of binary stars for the calculated synthetic CMD, given as a number between 0.0 and 1.0.
\item{$\bf{qmass}$}\\
Mass fraction for binary stars, given as a number between 0.0 and 1.0.  
\end{itemize}

\section{Examples}

A series of examples are now presented that illustrate the effect of varying the  input parameters one at a time.  Fig.~1 shows the reference synthetic CMD against which other CMD examples can be compared, 
The following input reference parameters were adopted: \\

\begin{center}
\noindent \bf{\underline{REFERENCE PARAMETERS}}\\
\noindent alpha = 0.\\
Z = 0.004\\
Age =8.0\\
N = 1000\\
IMF slope = 1.35\\
$M_{upper}$ = 5.\\
$M_{lower}$ = 0.6\\
ISEED =1807\\
Binary fraction = 0.\\
qmass = 0.5\\
\end{center}

\noindent Note that in every case below, only the one parameter discussed was varied; all other parameters are as given in the above reference parameter list.

\begin{itemize}
\item{$\bf{Varying~chemical~composition}$}\\

\subitem{$Metallicity~Z$}\\
Fig.~2 shows the effect of changing the metallicity Z from 0.004 to 0.02, all other input parameters being kept the same as in Fig.~1.  The reference heavy element solar mixture is the 
solar mixture of Grevesse \& Noels (1993).  It is given in Table~1, taken from Paper~2.

\subitem{$The~parameter~alpha$}\\
Fig.~3 shows the effect of varying $[\alpha/Fe]$ from 0. to 0.3.
Table~2, taken from Paper~2, lists the corresponding changes in total $Z$.   The reader is referred to Paper~2 for a discussion of the different effects of variations in $[\alpha/Fe]$  in low-$Z$ and high-$Z$ mixtures.  
\begin{deluxetable}{cccc}
\tablewidth{0pt}
\tablecaption{Conversion from [Fe/H] to $Z$ \label{tbl2}}
\tablehead{ \colhead{[Fe/H]}  &
\multicolumn{3}{c}{$Z$}\\ & \colhead{[$\alpha$/Fe]=0.0} &
 \colhead{[$\alpha$/Fe]=+0.3}  & \colhead{[$\alpha$/Fe]=+0.6} }
\startdata
-3.0 & 0.000019 & 0.000032 & 0.000058\\
-2.5 & 0.000062 & 0.000102 & 0.000182\\
-2.0 & 0.000195 & 0.000321 & 0.000574\\
-1.5 & 0.000615 & 0.001012 & 0.001807\\
-1.0 & 0.001935 & 0.003174 & 0.005627\\
-0.5 & 0.006021 & 0.009774 & 0.016990\\
 0.0 & 0.018120 & 0.028557 & 0.047000\\
 0.5 & 0.049711 & 0.072793 & 0.106471\\
 1.0 & 0.110798 & 0.142689 & 0.177489\\
\enddata
\end{deluxetable}

 \subitem{$Relative~H~and~He~abundances$}\\
Note that in the YY isochrone tables, the helium content by mass Y is kept fixed for a given value of $Z$.    In all examples given here, $(\Delta Y/\Delta Z)$ = 2 has been assumed.  This corresponds  to  $Y$ = 0.238 for $Z$ = 0.004, and to $Y$ = 0.27 for $Z$ = 0.02.

\begin{figure}
\plotone{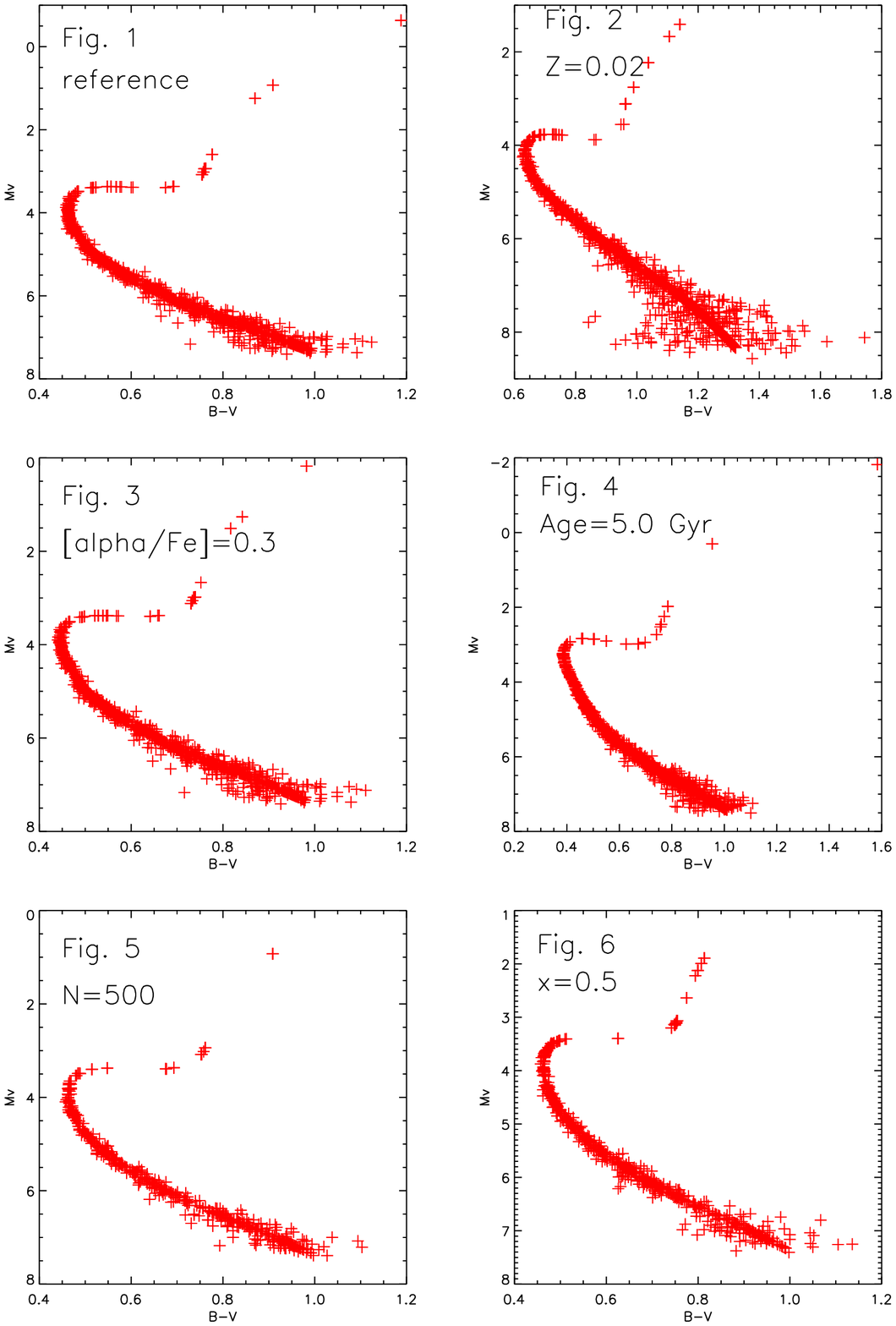}
\end{figure}

\begin{figure}
\plotone{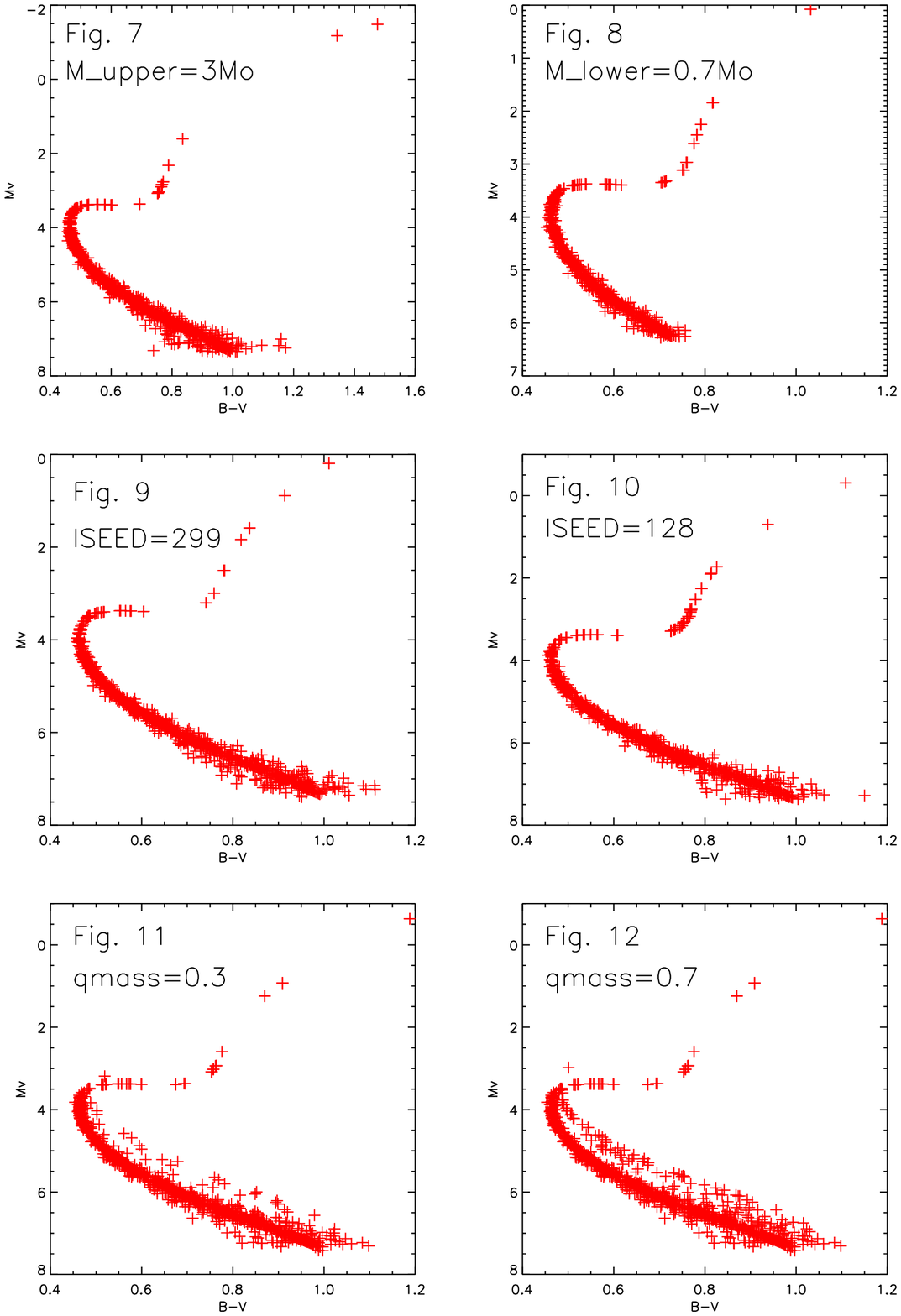}
\end{figure}

\item{$\bf{Varying~the~age}$}\\
In Fig.~4, the age has been changed from 8.0 to 5.0 Gyr.

\item{$\bf{Varying~the~total~number~of~stars~N}$}\\
In Fig.~5, the value of $N$ is changed from 1000 to 500.  As expected, stochastic fluctuations become larger as  $N$ decreases.

\item{$\bf{Varying~the~mass~function~parameters}$}\\
The distribution of stellar masses in the synthetic model depends on three parameters, the IMF slope 
$x$
and the upper and lower masses, $M_{upper}$ and $M_{lower}$ respectively.  The effect of changing the IMF slope from $x$ = 1.35 to 0.5 is shown in  Fig.~6.  The effects of changing independently the upper and lower mass limits to 3.0$M_{\odot}$ and 0.7$M_{\odot}$ are shown in Fig.~7 and Fig.~8, respectively, all other parameters are as given in the reference parameter list. 

\item{$\bf{ISEED}$}\\
Varying ISEED highlights the role of stochastic effects.  For small N, stochastic effects become 
significant in rapid evolutionary phases.  For example, the two CMDs shown in Fig.~9 and Fig.~10 are based on the same set of input parameters (except that ISEED=299 and 128, respectively).  This introduces an uncertainty in the integrated  energy distribution  of the stellar  population.

\item{$\bf{Varying~the~binary~star~content}$}\\
The presence of unresolved binaries can also affect the CMD significantly.
Fig.~1 shows a synthetic CMD that does not include binaries.  Fig.~11 and Fig.~12 are plotted for a binary fraction of 0.2, but with the different mass fractions qmass = 0.3 and 0.7, respectively. 
The synthetic cluster CMDs by Woo et al. (2000) illustrate the importance of including binary systems, particularly in moderately rich stellar systems. 

\end{itemize} 

\section{Conclusions}
This note describes a web-based CMD calculator designed 
for the study of stellar 
populations in the Galaxy and distant stellar systems.  This 
web based calculator, which is based on the YY isochrones, is a 
useful tool for studies of resolved star clusters, and for explorations 
of the sensitivity of the integrated light of stellar systems to 
stellar and population parameters.   

Future versions of this calculator will be based on a more comprehensive YY database 
covering an extended range of helium abundances, the inclusion of helium burning 
phases of evolution, often very significant in the CMDs of stellar systems, 
and the availability of a simple spectral energy distribution (SED) library.        

This research was supported in part by NASA grant HST-GO-10505.03-A.

\section{References}
\noindent Demarque, P., Woo, J.-H., Kim, Y. -C. \& Yi, S.  2004, ApJS, 155, 667 (Paper~4)\\
Green, E. M., Demarque, P., \& King, C. R. 1987, The Revised Yale Isochrones and Luminosity Functions (New Haven: Yale Univ. Obs.)\\ 
Grevesse, N., \& Noels, A. 1993, in Origin and Evolution of the Elements, ed. N. Prantzos, 
E. Vangioni-Flam, \& M. Cass\'{e} (Cambridge: Cambridge Univ. Press)\\
Kim, Y. -C., Demarque, P., Yi, S. \& Alexander, D.R.  2002, ApJS, 143, 499 (Paper~2)\\
Lejeune, Th., Cuisinier, F. \& Buser, R. 1998, A\&A, 130, 65\\
Park, J.-H. \& Lee, Y.-W. 1997, ApJ, 476, 28\\
Salpeter, E.E. 1955, ApJ, 121, 161\\
Tinsley, B.M. 1980, Fund. Cosmic Phys., 5, 287\\
Woo, J-H., Gallart, C., Demarque, P., Yi, S., Zoccali, M. 2003, AJ, 125, 754\\
Yi, S., Demarque, P.,  Kim, Y. -C.,  Lee, Y.-W., Ree, C.H. Lejeune, Th. \& Barnes, S. 2001, ApJS, 136, 417 (Paper~1)\\
Yi, S., Kim, Y -C., \& Demarque, P. 2003, ApJS, 144, 259 (Paper~3)\\

 \end{document}